\documentstyle[epsf]{astropart}
\renewcommand{\today}{18 August 1999}
\newcommand{\eqb}{\begin{eqnarray}}
\newcommand{\eqe}{\end{eqnarray}}
\newcommand{\gesim}{\,\raisebox{-0.4ex}{$\stackrel{>}{\scriptstyle\sim}$}\,}
\newcommand{\lesim}{\,\raisebox{-0.4ex}{$\stackrel{<}{\scriptstyle\sim}$}\,}
\newcommand{\PSR}{PSR~B1259$-$63}
\newcommand{\PSRjb}{PSR~J0045$-$73}
\newcommand{\diff}{{\rm d}}

\newcommand{\sigmaT}{\sigma_{\rm T}}

\newcommand{\Rstar}{R_*}
\newcommand{\Lstar}{L_*}

\newcommand{\betawind}{\beta_{\rm w}}
\newcommand{\gammawind}{\gamma_{\rm w}}
\newcommand{\rhowind}{\rho_{\rm w}}
\newcommand{\Lwind}{L_{\rm w}}
\newcommand{\epsoutplus}{\epsilon_{\rm out}^+}
\newcommand{\epsoutminus}{\epsilon_{\rm out}^-}
\newcommand{\epsout}{\epsilon_{\rm out}}
\newcommand{\epsin}{\epsilon_{\rm in}}

\newcommand{\epsinp}{\epsilon_{\rm in}'}
\newcommand{\thetaout}{\theta_{\rm out}}
\newcommand{\thetain}{\theta_{\rm in}}
\newcommand{\thetaoutp}{\theta_{\rm out}'}

\newcommand{\thetascattp}{\theta_{\rm sc}'}
\newcommand{\thetapulsar}{\theta_{\rm p}}
\newcommand{\me}{m_{\rm e}}
\newcommand{\Ein}{E_{\rm in}}
\newcommand{\Einp}{E_{\rm in}'}
\newcommand{\Eout}{E_{\rm out}}

\newcommand{\rT}{\hat{r}_T}
\begin{document}
\begin{frontmatter}
\title{Probing Pulsar Winds Using Inverse Compton Scattering}
\author[RCfTA]{Lewis Ball}
\and
\author[MPIK]{J. G. Kirk}
\address[RCfTA]{Research Centre for Theoretical Astrophysics,
University of Sydney, N.S.W. 2006, Australia}
\address[MPIK]{Max-Planck-Institut f\"ur Kernphysik,
Postfach 10 39 80, D-69029, Heidelberg, Germany}

\begin{abstract}
We investigate the effects of inverse Compton scattering by electrons
and positrons in the unshocked winds of rotationally-powered binary pulsars.
This process can scatter low energy target photons to produce
gamma rays with energies from MeV to TeV.
The binary radio pulsars \PSR\ and \PSRjb\ are both in close
eccentric orbits around bright main sequence stars which provide a
huge density of low energy target photons.
The inverse Compton scattering process transfers momentum from the
pulsar wind to the scattered photons, and therefore provides a drag which
tends to decelerate the pulsar wind.
We present detailed calculations of the dynamics of a pulsar wind which
is undergoing inverse Compton scattering, 
showing that the deceleration of the wind of \PSR\ due to
`inverse Compton drag' is small,
but that this process may confine the wind of \PSRjb\ before it attains
pressure balance with the outflow of its companion star.
We calculate the spectra and light curves of the resulting inverse Compton
emission from \PSR\ and show that if the size of the pulsar wind nebula is
comparable to the binary separation,
then the $\gamma$-ray emission from the unshocked wind may be
detectable by atmospheric Cerenkov detectors or by the new generation
of satellite-borne $\gamma$-ray detectors such as INTEGRAL and GLAST.
This mechanism may therefore provide a direct probe of the
freely-expanding regions of pulsar winds, previously thought to be invisible. 
\end{abstract}
\end{frontmatter}


\newpage

Rotationally-powered pulsars are rapidly rotating neutron stars whose rotation 
is slowing at a readily measurable rate.
It is generally accepted that the rotational energy is carried away by a 
wind of electrons, positrons, and possibly ions.
Models for the best studied rotationally-powered pulsar -- the Crab --
suggest that the wind expansion is highly relativistic with a bulk Lorentz 
factor of $\gammawind\sim 10^6$
[Rees \& Gunn 1974].
A number of authors have investigated the inverse Compton scattering of
emission from the neutron star itself by the
electrons and positrons of the pulsar wind both close to the surface
[Zhang, Qiao \& Han 1997; Qiao \& Lin 1998; Luo \& Protheroe 1998]
and at about the distance of the light-cylinder
[Bogovalov \& Aharonian 1999;
Aharonian \& Bogovalov 1999].
At much larger radii, pulsar winds may be confined by pressure balance
with the surrounding medium, as in the Crab nebula,
in which case the electrons and positrons in the wind will be
accelerated and isotropised at the shock which terminates the
relativistic wind.
The inverse Compton emission by electrons and
positrons in the shocked pulsar wind has been considered, addressing both
the scattering of self-produced synchrotron photons
[De Jager \& Harding 1992],
of diffuse background target photons 
[for a review see Harding 1996],
and of photons from a binary companion
[Kirk, Ball \& Skj\ae raasen 1999].
However, until recently, 
it has generally been thought that the intermediate region
between the light cylinder and the termination shock 
could not be observed directly.

In this paper we investigate inverse Compton scattering by the
electrons and positrons in this unshocked region of pulsar winds.
Chernyakova \& Illarionov [1999]
have also considered aspects of 
this problem, such as the deceleration of
the pulsar wind when the scatterings can be treated in the Thompson limit,
and the spectrum of scattered photons when the wind
deceleration and absorption by photon-photon pair production 
could be neglected.
Here we present a more detailed treatment, including deceleration in the 
Thomson and Klein-Nishina regimes and the effects of absorption.
We apply our formalism to 
two binary pulsar systems in which there is a huge density of
target photons from a luminous companion star and 
show that the pulsar \PSR\ should produce detectable gamma-ray
emission from the scattering of the companion star photons by
the unshocked pulsar wind,
provided it is not confined to a small region around the pulsar.
The strong dependence of the target photon density on the
position with respect to the star,
coupled with the fact that the target photons may be
assumed to all be propagating radially from the star,
and so are unidirectional at each point where a scattering occurs,
implies that the scattered emission will have a strong
dependence on orbital phase.
In close binary systems the scattering process can have a significant
influence on the dynamics of the unshocked wind,
and may provide the first direct probe of the
freely-expanding region of a pulsar wind.

\PSR\ is the only known galactic radio pulsar which is orbiting
a main-sequence companion.
It is in a close, highly eccentric orbit ($e\sim 0.87$) around
SS2883, a B2e star of radius $\Rstar\sim 6 R_\odot$
and luminosity $\Lstar\sim 8.8\times 10^3\,L_\odot$
[Johnston et al.\ 1992, 1994, 1996].
The system is thought to be $1.5\,$kpc from the Earth
and at periastron the pulsar is only $23\Rstar\approx 10^{11}\,{\rm m}$
from its companion.
The only other similar binary pulsar system, \PSRjb\
[Kaspi et al.\ 1994a,b; Bell et al.\ 1995],
is in the Small Magellanic Cloud, and is the most distant pulsar known.
It is in an orbit of eccentricity $0.81$ around an even more
luminous companion, a B1 V star of radius $6.4 R_\odot$ and
luminosity
$1.2\times 10^4\,L_\odot$,
and at periastron the pulsar is just $26R_\odot\approx 2\times10^{10}\,{\rm m}$
from its companion.

The spin-down luminosity of \PSR\ is significant,
$L_{\rm p} = 8.3\times 10^{28}\;$W,
and it is likely that this drives a relatively strong wind.
The wind is most probably confined by pressure
balance with the strong outflow from the pulsar's Be-star companion
[Melatos, Johnston \& Melrose 1995],
in which case the pulsar wind slows to a subsonic outflow
at a termination shock.
The already relativistic electrons and positrons from the pulsar wind
are isotropised and accelerated to even higher energies at this shock.
The unpulsed X-rays observed throughout the binary orbit
of this system
[Hirayama at al.\ 1996]
are thought to result from synchrotron emission from electrons
and positrons in the shocked region of the pulsar wind
[Tavani \& Arons 1997].
Kirk et al.\ [1999]
have shown that inverse Compton scattering of photons from the Be star,
by the electrons and positrons of the shocked pulsar wind,
is likely to produce a flux of hard $\gamma$-rays that would be
detectable by atmospheric Cerenkov telescopes.

If just $10^{-3}$ of the spin down luminosity of \PSR\ is
converted into $\gamma$-rays by inverse Compton scattering in
the {\em unshocked} wind, the resulting flux would be detectable
at energies somewhere in the MeV--TeV range depending on the
Lorentz factor of the pulsar wind.
The lower spin down luminosity and greater distance to \PSRjb\ means
that it should be unobservable in hard $\gamma$-rays.
Nevertheless the binary separation of this system is even smaller than that
of \PSR\ and the companion star is also more luminous,
so inverse Compton scattering by the unshocked wind may 
affect the dynamics of this system.

In the following sections we present equations
describing the deceleration of the pulsar
wind by `inverse Compton drag'
-- the energy losses of the unshocked electrons and positrons
due to inverse Compton scattering of the companion-star photons,
and the intensity and spectrum of the energetic scattered photons.
We show that the inverse Compton losses cannot contain the wind of \PSR\
before it attains the radius at which pressure balance with the
Be-star outflow occurs.
These losses may significantly decelerate
the wind of \PSRjb\ at times near periastron within a range of angles about
the line joining the pulsar and its companion.
We present examples of the emitted inverse-Compton spectrum and of
the light curves from \PSR , showing that the emission from
the unshocked wind should be detectable as gamma-rays.
The signal from the unshocked wind is quite different from that from
the shocked wind, so that if a variable gamma-ray signal is detected from
this system, these two possible source mechanisms should be
readily distinguishable.

\section{Dynamics of the pulsar wind}
We assume that a relativistic, radially directed wind containing electrons,
positrons and an admixture of ions emerges from the pulsar.
This wind propagates in an ambient radiation field and the particles in the
wind interact incoherently with the ambient photons.
The most important interaction is Compton scattering by the electrons and
positrons.
In this section we outline the essential steps in the derivation of the 
equations for the dynamics of a pulsar wind subject to
inverse Compton scattering.
Further details are presented in Appendix A. 

Observations and models of the wind of the Crab pulsar suggest that it 
emerges nearly tangentially at around the light cylinder radius
and that the Poynting flux dominates the outflow at such radii
[Rees \& Gunn 1974].
However, by the time the wind interacts with the surrounding nebula
the importance of the magnetic field is greatly diminished
and the outgoing flux is dominated by the particle momentum.
Models suggest that the wind ultimately attains a bulk Lorentz factor of 
order $\gammawind=10^6$
[Kennel \& Coroniti 1984; Emmering \& Chevalier 1987].
While this picture is well accepted, the physics responsible for
the transformation of the wind is poorly understood.
Nevertheless, it is likely that the wind evolves relatively rapidly
into a cold, relativistic, particle dominated, radial outflow.
We will assume that at the radii of interest here,
which are well beyond the light cylinder, the electrons and positrons of
the pulsar wind may be taken to be monoenergetic, with Lorentz factor 
equal to that of the bulk flow, $\gammawind$.

We denote by 
$n_\gamma(\epsilon,{\bf R},\vec{\Omega}) \,\diff\vec{\Omega} \, \diff \epsilon$
the differential number density of target photons moving within the solid angle
$\diff\vec{\Omega}$ of the unit vector $\vec{\Omega}$ at position
${\bf R}$ with respect to the star,
with energy between $E=\epsilon \me c^2$ and $(\epsilon+\diff\epsilon)\me c^2$.
We investigate the deceleration of the wind in
the highly anisotropic radiation field comprising the light from a companion star
which is treated as a point source.
The target photons are treated as monochromatic. 

\newpage

An individual electron of Lorentz factor $\gamma$ scatters 
target photons at an average rate
\eqb
\dot{N}_{\rm ic}&=& 
c\int \diff\vec{\Omega}\int\diff{\epsin}\;
{{\epsinp}\over\gamma{\epsin}}\;
n_\gamma({\epsin},{\bf R},\vec{\Omega})
\sigma_{\rm KN}({\epsinp})
\label{scattrate}
\eqe
[Jones 1965],
where $\sigma_{\rm KN}(x)$ is the Klein-Nishina cross section.
(Note that when used alone the term `electron' may be taken to refer to 
both electrons and positrons in the pulsar wind.)
Unprimed quantities are measured in the observer's frame and primed
quantities are measured in the rest frame of the incoming electron.
Energies are expressed in units of $\me c^2$ so that 
$\Ein={\epsin}\me c^2$ is the energy of the target photon in the
observer's frame, 
$\Einp={\epsinp}\me c^2$ the energy of the target photon seen in the
electron rest frame.
The process of triplet pair production, ${\rm e}\gamma\rightarrow 3{\rm e}$,
can be ignored since the maximum value
of $\gamma\epsin$ which we consider is of the order 100
[Mastichiadis 1991].

The average rate of change of energy of the 
photon field is given by an integral similar to that in
Eq.~(\ref{scattrate}),
but weighted by the factor ${\epsout}-{\epsin}$,
where $\Eout=\epsout \me c^2$ is the energy of the scattered photon given by 
\eqb
{\epsout}&=&
{{\epsinp}\gamma(1-\beta\cos\thetaoutp)\over
1+{\epsinp}(1-\cos\thetascattp)}
\eqe
with $\thetaoutp$ the angle between the outgoing photon and the 
velocity vector of the observer's frame and $\thetascattp$ the scattering angle,
both measured in the electron rest frame.
The rate of change of radial momentum is calculated in a similar fashion,
using $-{\epsout}\cos\thetaout+{\epsin}\cos\thetain$ as the weighting function,
with $\thetain$ and $\thetaout$ the angles between the incoming and outgoing
photons and the vector opposite the incoming electron velocity in the
observer's frame.

As a result of Compton scattering the electrons and positrons
of the pulsar wind lose both energy and radial momentum,
but gain a small amount of transverse momentum,
so that the gas is both decelerated and heated.
In the case of an ultra-relativistic wind the heating is small
and may be neglected (see Appendix A).
The deceleration of the wind due to inverse Compton drag is then given by the
equation
\eqb
{\diff\gammawind \over\diff r} 
&=& -{3\eta\sigmaT\over8\betawind}\int\diff\vec{\Omega}\diff{\epsin} 
{{{\epsinp}}^2\over{\epsin}}
\left(1-{{\epsin}\over\gamma{\epsinp}}-{{\epsin}\over\gamma}\right)
n_\gamma({\epsin},{\bf R},\vec{\Omega}) F_{\rm loss}({\epsinp})
\label{energy2}
\eqe
in which $\betawind=(1-\gammawind^{-2})^{1/2}$,
$\sigmaT$ is the Thompson cross section,
and $F_{\rm loss}$ is a function defined by Jones~[1965]
\eqb
F_{\rm loss}(x)&=&
{-2x(10x^4-51x^3-93x^2-51x-9)
\over
{3x^4(1+2x)^3}}
\nonumber\\
&&+{(x^2-2x-3)\ln(2x+1) \over x^4}
\label{Flossa1}
\eqe
[see also 
Kirk et al.~1999].
The parameter $\eta$ is the fraction of the 
wind luminosity carried by electrons and positrons: 
\eqb
\eta&=& {1\over 1+Mn_{\rm i}/(\me n_{\rm e}) }
\eqe
with $M$ the ion mass and $\me$ the electron mass,
and $n_{\rm i}$ and $n_{\rm e}$
the ion and electron number densities at any given radius.

\begin{figure}[htb]
\vspace*{0.3cm}
\epsfxsize=12 cm
\epsffile{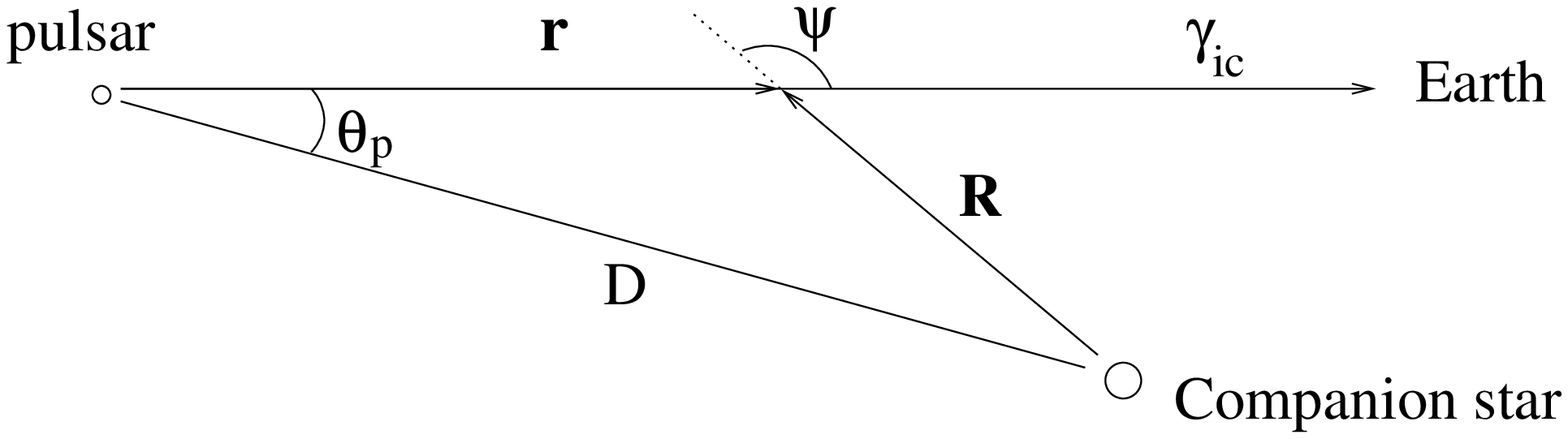}
\caption{Sketch of the binary system defining angles
and distances used in the text.}
\label{geometry}
\end{figure}

When the target photons originate from a companion star, 
the target field can be written in the form
\eqb
n_\gamma({\epsin},{\bf R},\vec{\Omega})&=&
N(R)\delta({\epsin}-\epsilon_0)\delta(\vec{\Omega}-\hat{{\bf R}})
\label{binaryrad}
\eqe
where $N(R)=L_*/(4\pi R^2ch\nu_0)$ is the photon density, 
$\epsilon_0=h\nu_0/\me c^2$ is 
the dimensionless energy of the companion star photons, 
$\hat{{\bf R}}={\bf R}/R$ is a unit vector, and $L_*$ is the stellar luminosity. 
We denote the binary separation of the pulsar and its companion star by $D$,
and introduce the dimensionless distance $\hat r=r/D$.
Inserting Eq.~(\ref{binaryrad}) into (\ref{energy2}), making the 
approximation $\betawind\approx1$, we find
\eqb
{\diff\gammawind
\over\diff \hat r} 
&\approx&
-{3\eta N_0\sigmaT D\over8\sin^2\thetapulsar}\;
{{{\epsinp}}^2\sin^2\psi\over\epsilon_0}
\left(1-{\epsilon_0\over\gammawind{\epsinp}}-{\epsilon_0\over\gammawind}\right) 
F_{\rm loss}({\epsinp})
\label{compdrag}
\eqe
where $N_0=N(D)$, ${\epsinp}=\gammawind\epsilon_0(1-\betawind\cos\psi)$ and 
$\psi$ is the angle between the vectors ${\bf R}$ and ${\bf r}$.
The geometry is shown in Fig.~\ref{geometry}.
Note that the angle $\psi$ is a function of position $\hat r$:
\eqb
\cos\psi&=&{\hat r -\cos\thetapulsar\over
\sqrt{1+{\hat r}^2-2\hat r\cos\thetapulsar}}\;.
\eqe

Equation (\ref{compdrag}) implies that $\dot \gammawind>0$ when
$\epsinp < \epsilon_0/(\gammawind - \epsilon_0)$
meaning that the target photons actually accelerate electrons
traveling almost directly away from the companion star
for which $\cos{\psi}\approx 1$.
The condition for this to occur is
\eqb
\cos{\psi}> {1\over \sqrt{1-1/\gammawind^2}} \;
            \left( 1 - {1\over {\gammawind(\gammawind-\epsilon_0)}}\right)\;;
\eqe
for $\gammawind=10^6$ and $\epsilon_0=10^{-5}$ this
is equivalent to $|\psi|\lesim 10^{-6}$, which is satisfied only
when the electrons are moving essentially radially away from the companion star.

\section{An approximate treatment \label{approx}}
Before applying the results of \S 1 and deriving the spectrum of the
scattered radiation, we consider a greatly simplified treatment obtained by
approximating the scattering cross-section by the Thompson cross-section
and assuming that the energy distribution of the scattered photons is a delta function.
In this approximation an electron of Lorentz factor $\gamma$ emits monochromatic 
photons of energy $\epsout=4\gamma^2\epsin/3$ and the reaction rate is simply 
$N \sigmaT c$. 
The equation governing the deceleration of the wind
is then
\eqb
{\diff\gammawind\over\diff r}&\approx&
-{4\over3}\eta\gammawind^2\epsin N \sigmaT\;.
\eqe
If the density of target photons remains roughly constant on
the length scale upon which the wind is decelerated, it follows that 
\eqb
r_{\rm decel}&\approx&
\left[\eta\gammawind(0)\epsin N \sigmaT\right]^{-1}
\\
&=&7.8\times 10^{9}
\left({10^6\over\eta\gammawind(0)}\right) \left({10^{12} {\rm eV\,cm^{-3}}
\over U_{\rm rad}}\right)
{\rm \, meters},
\label{lscaleic}
\eqe
where the scaling of the energy density of the target photons
has been chosen to be close to that of \PSR\ at periastron.

We next define a dimensionless spectral energy distribution ${\cal F}(r)$,
which describes how the radiation builds up along the line of sight within the 
pulsar wind.
At the outer boundary $r=r_{\rm max}$ of the source, the final, observable 
value is attained:
${\cal F}(r_{\rm max})=EF_E4\pi d^2/\Lwind$
where $F_E$ is the energy flux density at the Earth of photons of energy $E$,
and $d$ is the distance to the pulsar.
The equation for the generation of the spectral energy density is then
\eqb
{\diff{\cal F}\over\diff{r}}
&=&
{\eta\over\gammawind(0)}\;N \sigmaT \epsout^2\delta
\left(\epsout-4\gammawind^2\epsin/3\right)\;.
\eqe
Under the same assumption that the density of target photons
is roughly constant over the distance $r_{\rm decel}$,
 the emitted spectral energy distribution is
\eqb
{\cal F}&=&{1\over2}\sqrt{\epsout\over\epsoutplus}
\label{spectrumic}
\eqe
extending from a maximum at $\epsout=\epsoutplus=4\gammawind^2(0)\epsin/3$
down to a minimum which is a function
of the physical extent of the wind.

The approximations given by Eqs.~(\ref{lscaleic}) and (\ref{spectrumic})
can be applied to \PSR\ and \PSRjb\ at periastron using
the appropriate physical parameters which are summarized in Table 1.
The energy density in the target radiation field near the surface
of \PSR\ at the time of periastron is roughly
$6\times10^{11}\,{\rm eV\,cm^{-3}}$, 
which implies $r_{\rm decel}\approx 1.3\times 10^{10}\,{\rm m}\sim 0.1\,D_\tau$,
where $D_\tau=9.6\times10^{10}\,{\rm m}$ is the separation of the stars at periastron.
Taken at face value this estimate suggests that the inverse Compton drag 
may be so efficient as to decelerate the pulsar wind completely
before it attains the radius at which it would otherwise reach pressure
balance with the Be star outflow.
Furthermore this estimate implies that there should be a substantial energy flux at the Earth
($\sim 10^3\,{\rm eV\,cm^{-2}\,s^{-1}}$) concentrated in TeV photons.
This corresponds to a photon flux of $\sim 10^{-9}\,{\rm cm^{-2}\,s^{-1}}$,
roughly $10^3$ times higher than the sensitivity threshold
of the imaging Cerenkov detector operated by the CANGAROO collaboration,
and clearly conflicts with the marginal detection found by that experiment
[Sako et al.\ 1997].

For \PSRjb\ the energy density in the target radiation field at periastron
is even higher, $U_{\rm rad}\sim2\times10^{13}\,{\rm eV\,cm^{-3}}$,
whence $r_{\rm decel}\approx 1.3\times 10^{10}\,{\rm m}\sim 0.02\,D_\tau$
with $D_\tau=1.8\times10^{10}\,{\rm m}$ for this system.
However, the much greater distance of this source makes it uninteresting for 
gamma-ray observations.

These simple estimates indicate that inverse Compton drag on the unshocked winds
of these binary pulsars may be important.
However the Klein-Nishina corrections to the scattering cross section are
significant in these systems,
as are the effects of the changing angle $\psi$ and the changing
density of target photons encountered by the scattering electrons
as they propagate radially from the pulsar.
We include these effects in the next sections.

\newpage

\begin{table}[h]
\caption{Parameters for the \PSR\ and \PSRjb\ systems.
References are:
1--Johnston et al.\ [1996];
2--Johnston et al.\ [1994];
3--Underhill \& Doazan [1982];
4--Kaspi et al.\ [1994a];
5--Kaspi et al.\ [1994b];
6--Bell et al.\ [1995].}
\label{param}
\begin{tabular}{llcccc}
{\bf Parameter}         & ~             & \multicolumn{2}{c}{\bf \PSR} & \multicolumn{2}{c}{\bf \PSRjb} \\
                        &               & {\bf Value}                       & {\bf Ref.} & {\bf Value}                       & {\bf Ref.} \\
{\bf Pulsar}\\
period                  & $P$           & $47.762 \;{\rm ms}$               & 1          & $ 0.926 \;{\rm s}$                & 4 \\
period derivative       & $\dot{P}$     & $2.279\times 10^{-15}$            & 1          & $4.486\times 10^{-15}$            & 5 \\
surface magnetic field  & $B_{\rm p}$   & $3.3\times 10^7\;{\rm T}$         & 1          & $2.1\times 10^8\;{\rm T}$         & 4 \\
spin down luminosity    & $L_{\rm p}$   & $8.3\times 10^{28}\;{\rm W}$      & 1          & $2.2\times 10^{25}\;{\rm W}$      \\
{\bf Companion star}\\
spectral type           &               & B2                                & 2          & B1 V                              & 6 \\
effective temperature   & $T_{\rm eff}$ & $2.28\times 10^4\; {\rm K}$       & 3          & $2.4\times 10^4\; {\rm K}$        & 6 \\
radius                  & $R_*$         & $6.0 R_\odot$                     & 3          & $6.4 R_\odot$                     & 6 \\
                        &               & ($4.2\times10^9\;{\rm m}$)        &            & ($4.5\times10^9\;{\rm m}$)        \\
luminosity              & $L_*$         & $8.8\times 10^3 L_\odot$          & 3          & $1.2\times 10^4 L_\odot$          & 6 \\
                        &               & ($3.3\times10^{30}\;{\rm W}$)     &            & ($4.6\times10^{30}\;{\rm W}$)     \\
effective photon energy & \multicolumn{2}{l}{$\epsilon_0 = 2.7k_{\rm B}T_{\rm eff}/(m c^2)$} \\
                        &               & $10^{-5}$                         &            & $10^{-5}$                         \\
mass                    & $M_*$         & $10 M_\odot$                      & 3          & $8.8 M_\odot$                     & 6 \\
                        &               & ($2\times10^{31}\;{\rm kg}$)      &            & ($1.8\times10^{31}\;{\rm kg}$)    \\
{\bf System}\\
eccentricity            & $e$           & 0.87                              & 1          & 0.81                              & 4 \\
periastron separation   & $D_\tau$      & $23 \Rstar$                       &            & $4 \Rstar$                        \\
                        &               & ($9.6\times10^{10}\;{\rm m}$)     &            & ($1.8\times10^{10}\;{\rm m}$)     \\
apastron separation     & $D_a$         & $331 \Rstar$                      &            & $38 \Rstar$                       \\
                        &               & ($1.4\times10^{12}\;{\rm m}$)     &            & ($1.7\times10^{11}\;{\rm m}$)     \\
orbital inclination     & $i$           & $35^\circ$                        & 1          & $44^\circ$                        & 6 \\
distance                & $d$           & $1.5\,$kpc                        & 1          & $57.5\,$kpc                       & 4 \\
\end{tabular}
\end{table}

\section{Inverse Compton emission from the unshocked wind}
The basic approximation made in computing the inverse Compton emission of a 
relativistic particle is that the scattered photons are emitted in the
direction of motion of the particle.
Thus in the case of scattering by a radial pulsar wind,
an observer sees photons produced at all points along the radius vector
from the pulsar to the point at which the wind is either decelerated
by radiation drag, 
or shocked, or simply runs out of target photons.
In addition, scattered photons of sufficiently high energy
may be absorbed by pair production on the same population of target photons
via the process $\gamma\gamma\rightarrow e^+e^-$.
The probability for pair production, like the scattering rate,
is a function of position along the radius vector.
In this section we derive expressions
for the inverse Compton emission from the unshocked wind,
including the effects of pair-production absorption.
Further details are presented in Appendix B.

Observations of the Crab nebula suggest that in that case at least,
the pulsar wind has an energy-dependent anisotropy
[Aschenbach \& Brinkmann 1975; Hester et al.\ 1995].
However, for simplicity we assume that the pulsar wind is
independent of latitude and longitude.
The distribution function of monoenergetic, radially directed electrons
in the wind may then be written as
\eqb
n_{\rm e}(\gamma,{\bf r},\vec{\Omega})&=&
N_{\rm e}(r)\delta\left[\gamma-\gammawind(r)\right]
\delta(\vec{\Omega}-\hat{{\bf r}}),
\label{windees}
\eqe
where $N_{\rm e}(r) = \eta\Lwind /\left[4\pi r^2
c\betawind\gammawind(0)\me c^2\right]$,
and $\Lwind$ is the luminosity of the pulsar wind,
which is taken to be equal to the spin-down luminosity.
It is relatively straightforward to incorporate a more complicated model,
provided the flow remains radial.

The specific intensity of radiation $I({\epsout},r)$ is governed by the 
radiative transport equation
\eqb
{\diff I\over\diff r}&=&
\left({\Lwind\over4\pi}\right){\eta\epsout\over \gammawind(0)\me c^3} 
\; {\diff N_{\gamma}\over\diff{\epsout}\diff t}
 + I \; {\diff\tau\over\diff r},
\eqe
where $\diff N_{\gamma}/\diff{\epsout}\diff t$ 
is the differential rate of emission of inverse Compton scattered photons
by a single electron, and
$\tau$ is the optical depth between the observer and the point 
$r$ due to pair production.
Since $r$ is measured from the pulsar outwards,
it follows that $\diff\tau/\diff r\le0$. 
More details are presented in Appendix B.

When the target radiation originates from a companion star, as specified in 
Eq.~(\ref{binaryrad}), the photon production rate can be written as
\eqb
{\diff N_\gamma\over\diff t\diff 
{\epsout}}&=&N(R)\sigmaT c \;{\cal H}({\epsout},\cos\psi),
\label{photonproduction}
\eqe
where ${\cal H}({\epsout},\cos\psi)$ is related to the quantity
${\diff N_\gamma/\diff t\diff{\epsout}[\diff{\epsin}\diff x']}$ derived by
Ho \& Epstein~[1989]
and is defined in Appendix B.

In terms of the dimensionless radius $\hat r=r/D$, the 
dimensionless spectral energy distribution ${\cal F}(r)$
introduced in the previous section 
(${\cal F}(\infty)=4\pi d^2 EF_E/\Lwind$
where $F_E$ is the flux and $d$ the distance to the object)
builds up in the pulsar wind according to:
\eqb
{\diff {\cal F}\over \diff \hat r}&=& 
\left({\eta N_0\sigmaT D \over\gammawind(0)\sin^2\thetapulsar}\right)
{\epsout}^2{\cal H}({\epsout},\cos\psi)\sin^2\psi
+{\diff\tau\over\diff\hat r}{\cal F},
\label{transport}
\eqe
where $\thetapulsar$ is the angle between 
the line joining the stars and the line of sight. 
The absorption optical depth obeys the equation
[Kirk et al.~1999]
\eqb
{\diff\tau\over\diff\hat r}&=&
-\left({N_0\sigmaT D\over\sin^2\thetapulsar}\right)
\sin^2\psi(1-\cos\psi)\;\hat{\sigma}_{\gamma\gamma}(\bar\epsilon),
\label{opacity}
\eqe
where $\hat{\sigma}_{\gamma\gamma}(z)$ is the cross section for pair production 
in units of $\sigmaT$ (see Appendix B), which is to be evaluated with argument
$\bar\epsilon=\sqrt{{\epsout}\epsilon_0(1-\cos\psi)/2}$.

The computation of the inverse Compton luminosity of the pulsar wind in 
the radiation field of its companion star therefore reduces to the integration of 
equations (\ref{compdrag}) and (\ref{transport}), with the optical depth gradient 
given by (\ref{opacity}).

\section{Inverse Compton drag on the winds of\protect{\newline} \PSR\ \& \PSRjb}
\begin{figure}
\epsfxsize=14.cm\epsffile{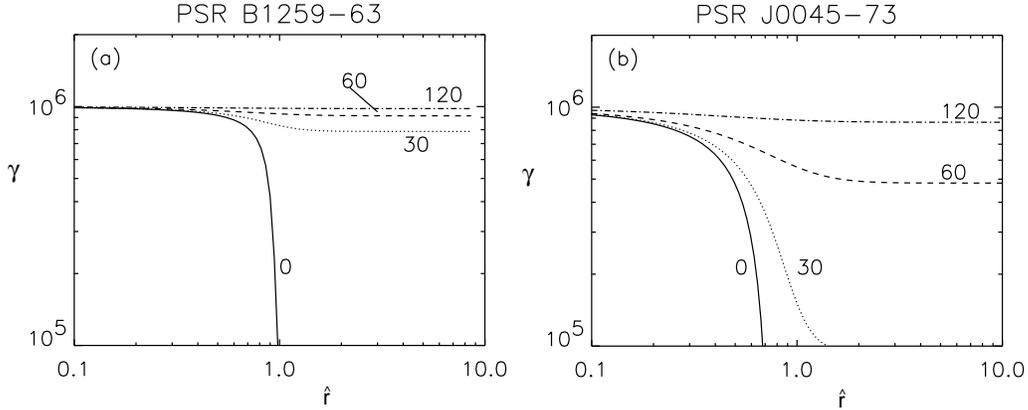}
\caption{Lorentz factor $\gammawind(\hat{r})$ of a radial
wind with $\gammawind(0)=10^6$ and $\eta=1$,
subject to inverse Compton drag from the
companion star photons, for \PSR\ (a) and \PSRjb\ (b).
The pulsar--companion star separation is that at periastron.
The curves correspond to different values
of $\thetapulsar$ (see Fig.~\protect\ref{geometry}) as labelled.}
\label{gammar}
\end{figure}

Equation (\ref{compdrag}) can be numerically integrated to obtain
the Lorentz factor of the radial wind as a function of $r$.
Figure \ref{gammar} shows the results for four different values of $\thetapulsar$,
for parameters appropriate to \PSR\ and \PSRjb\
as given in Table~\ref{param},
assuming that the wind momentum is carried entirely by electrons and 
positrons, i.e.\ $\eta=1$.
When $\thetapulsar=0$ the electrons are traveling straight towards the companion star
and inverse Compton drag is at its most effective because the
collisions are head on and the density of
target photons increases as $(1-\hat{r})^{-2}$.
For large values of $\thetapulsar$ the drag is relatively ineffective
because the electron--photon collisions are far from head on,
and furthermore, the target photon density drops as the electrons
propagate away from the companion star.

The curves for $\thetapulsar=0$ shown in Fig.~\ref{gammar} can be compared to
the simple estimates presented in \S \ref{approx}.
Figure \ref{gammar}a shows that in the head-on case, inverse Compton drag
stops electrons in the wind of \PSR\ at $\hat r\approx 0.9$
compare to the simple estimate $r_{\rm decel} \sim 0.1 D_\tau$. 
On the other hand, Fig.~\ref{gammar}b shows that head-on scatterings
stop the electrons in the wind of \PSRjb\ at $\hat{r}\approx0.7$,
compared to the simple estimate $r_{\rm decel} \sim 0.02 D_\tau$.
These differences arise primarily because the simple estimates
ignored the Klein-Nishina effects --- which reduce the scattering
cross section from the Thompson value.
The increasing density of target photons encountered by the
wind electrons as they approach the companion star offsets the
Klein-Nishina correction to some degree, but the net effect is that the
deceleration length is much larger than suggested by the simple estimate.

\begin{figure}
\epsfxsize=12cm\epsffile{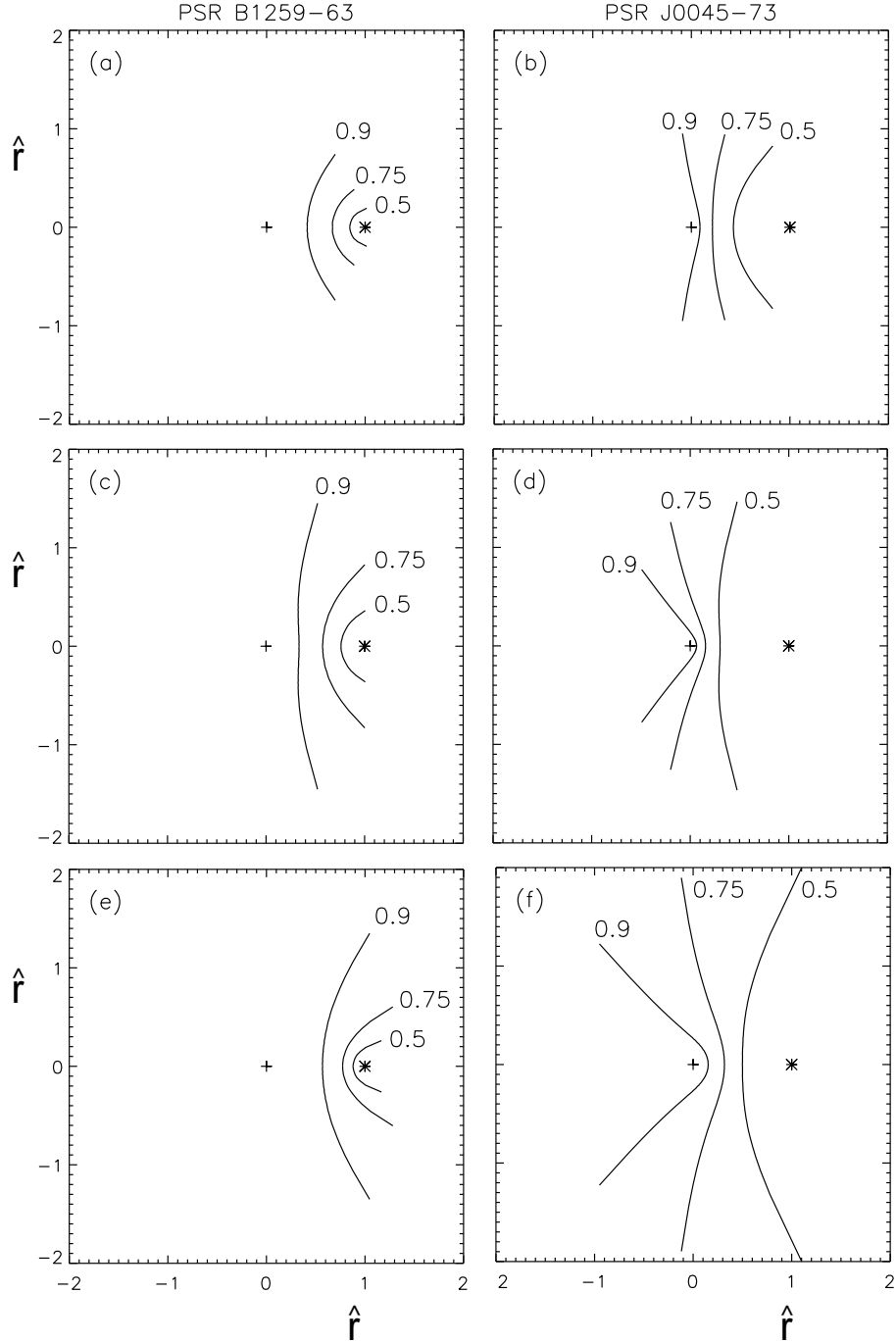}
\caption{Cross sections through the surfaces where $\gamma(\hat{r})$
has dropped to 90\%, 75\% and 50\% of the initial value $\gammawind(0)$
-- as labelled, if inverse Compton drag is the dominant wind
deceleration mechanism.
The wind, orbit and target photon parameters are as for
\PSR\ (left) and for \PSRjb\ (right)
at periastron and are given in Table~\protect\ref{param}.
The pulsar is at the origin and is marked by a plus
and the companion star is marked by an asterisk.
The two top panels (a) \& (b) show the results when $\gammawind(0)=10^4$,
the middle two (c) \& (d) are for $\gammawind(0)=10^5$,
and the lower two panels (e) \& (f) are calculated for $\gammawind(0)=10^6$.}
\label{shockgeom}
\end{figure}

Equation (\ref{compdrag}) can also be used to determine the position 
of the surface at which the radial wind loses a given fraction of its energy as
a result of inverse Compton scattering.
Such a surface is rotationally symmetric about the line joining the two stars.
Cross sections of the surfaces where $\gammawind(\hat{r})$ has fallen to
90\%, 75\% and 50\% of its initial value are shown in Fig.~\ref{shockgeom}
for \PSR\ and \PSRjb\ at periastron,
for three different values of the initial Lorentz factor $\gammawind(0)$.
Unless the interaction between the stellar wind and the pulsar wind
intervenes, we expect the pulsar wind to terminate roughly at the position
where the wind electrons have lost half of their initial energy.

The shapes of the cross-sections in all the panels of Fig.~\ref{shockgeom}
clearly show the strong dependence of the wind deceleration on the angle
$\thetapulsar$.
The surfaces of equal loss are all closest to the pulsar along the
line joining the pulsar to the Be star, and rapidly become more distant from the
pulsar as $\thetapulsar$ increases.
This occurs because while head-on collisions are effective at
decelerating the electrons,
the deceleration drops off very rapidly as $\thetapulsar$ increases,
and the photon density also drops off very rapidly with distance
from the pulsar at even moderate values of $\thetapulsar$.

The dependence on $\gammawind(0)$ as depicted in Fig.~\ref{shockgeom}
is particularly striking.
The surfaces of equal loss are much closer to the pulsar,
for both \PSR\ and \PSRjb ,
when $\gammawind(0)=10^5$ than for either $\gammawind(0)=10^4$ or $\gammawind(0)=10^6$.
This in an indication that the deceleration is most effective when
the system is `tuned' such that $\gamma\epsilon_0\sim1$.
This occurs because the energy loss rate is made up of two parts,
the scattering rate and the average energy loss per collision.
The former is constant in the Thomson regime, but falls off with increasing energy
in the Klein-Nishina regime.
The average energy loss per collision divided by the energy of the scattering particle
increases proportional to $\gamma$ in the Thomson regime,
and tends to a constant in the Klein-Nishina regime.
The combined result of these effects is that there is a peak in the energy loss rate
near the transition between the Thomson and Klein-Nishina regimes.

Finally, Fig.~\ref{shockgeom} shows the difference between
the inverse Compton deceleration the winds of \PSR\ and of \PSRjb.
In \PSR\ the density of photons from the companion star is not large enough to
substantially decelerate the pulsar wind before it has travelled a substantial
fraction of $D_\tau$, even along the line joining the stars where the collisions
are precisely head on.
The surfaces depicted, which all have their apexes at $\hat{r}\gesim0.25$,
therefore wrap around the Be star rather than the pulsar.
These results suggest that in \PSR\ the inverse Compton drag is unlikely
to terminate the pulsar wind before it attains pressure balance with the Be star outflow,
given that this balance is expected to occur at $\hat{r}\lesim0.5$
[Melatos et al.\ 1995].
It is therefore reasonable to assume, with
Kirk et al.\ [1999],
that a termination shock forms in the pulsar wind as a result of
pressure balance with the Be-star outflow.
Nevertheless, in the case of \PSR\ the conversion of even a
small fraction of the wind energy could result in an observable flux. 
From Fig.~\ref{shockgeom} it is clear that if the wind terminates at
a radius $\rT$ which is a significant fraction of the binary separation,
it may become observable.
For $\gammawind(0)\sim10^5$ and $\rT\sim0.5$ the figure indicates that
at periastron the unshocked wind will lose approximately
25\% of its momentum due to inverse Compton scattering along the line
joining the stars, resulting in a substantial gamma-ray luminosity.

For \PSRjb\ the situation is even more extreme.
The companion star photon density at the pulsar is so large at periastron
-- because the orbital separation is so small --
that inverse Compton drag is much more effective at decelerating the pulsar wind.
In this system many of the surfaces of equal loss depicted have their apex
very close to the pulsar and therefore wrap around the pulsar rather than
the companion star.
It is possible that inverse Compton drag may stop the wind of \PSRjb\
before it attains pressure balance with the companion star outflow, at least at angles
close to the line connecting the stars, and at times close to periastron.

\section{Emitted gamma rays}
%
\subsection{Spectra}
This section details the spectrum of inverse Compton emission produced via
scattering from the unshocked wind.
We assume that the wind momentum is carried entirely by electrons and 
positrons ($\eta=1$),
that the only deceleration is that from the scattering,
and that the wind expands freely without being terminated via pressure balance
with the companion star outflow ($\rT \gg 1$).
The spectra obtained by integrating equations
(\ref{compdrag}) and (\ref{transport}) with (\ref{opacity}),
for a range of wind Lorentz factors $\gammawind(0)$ and at three different angles
$\thetapulsar$ with other parameters relevant to \PSR\ at periastron,
are presented in Fig.~\ref{spectra}.
Note that the spectra show the energy flux versus the photon energy.
The emitted spectra depend dramatically on both the angle
$\thetapulsar$ and on $\gammawind(0)$.

For $\thetapulsar=5^\circ$ (a value not, however, observable from the Earth)
the wind electrons are initially directed almost
along the line connecting the pulsar and its companion.
Inverse Compton scattering is thus intrinsically efficient as the
electron-photon collisions are close to head on, and the electrons in the
wind encounter a very large target photon density as they pass close to
the companion star.
Integration over the scatterings occurring along such a radius therefore
produces a relatively broad spectrum of emitted photons because of the
wide range of Lorentz factors of the scattering electrons.
Furthermore, the scattered emission is relatively bright because the wind
loses a substantial fraction of its momentum to the scattered photons.

For $\thetapulsar=125^\circ$ the wind electrons are initially directed more
or less away from the companion star, so the scatterings are very far from
head on and the target photon density decreases very rapidly as the electrons
propagate away from the pulsar (and hence from the companion star).
The emitted spectrum in this case is close to monochromatic because the
electron deceleration is very small so the scattering particles are
essentially monoenergetic.
Furthermore, since only a small fraction of the wind momentum is
transferred to the scattered photons, the flux of the scattered photons
is commensurately small.

\clearpage

\thispagestyle{empty}

\begin{figure}
\epsfxsize=14cm\epsffile{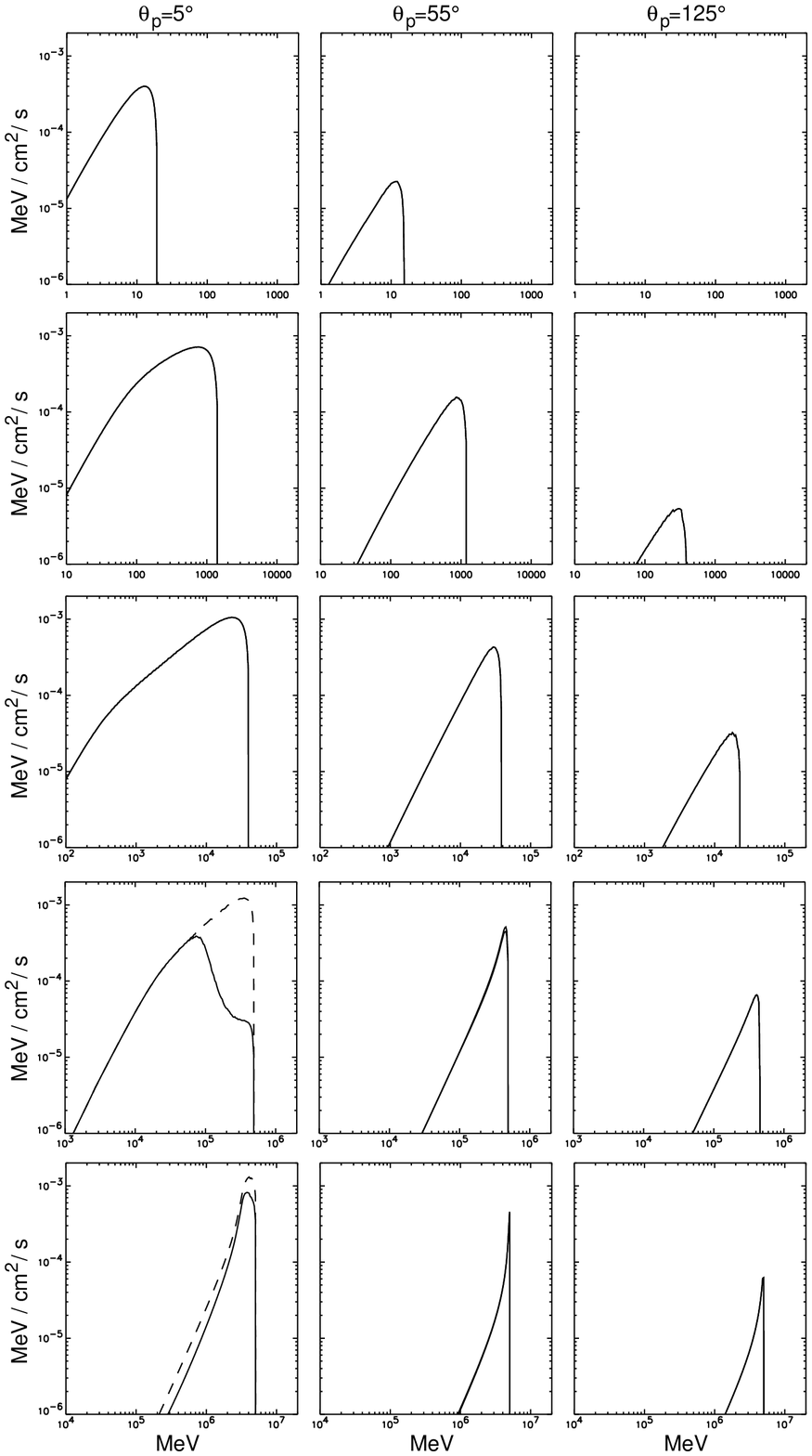}
\end{figure}

\clearpage

\begin{figure}
\caption{The SED (spectral energy distribution) $EF_E$ 
with $F_E$ the energy flux at the distance of the Earth of
inverse Compton scattered photons as a function of photon energy -- at
angles $\thetapulsar=5^\circ$ (left), $65^\circ$ (middle),
and $125^\circ$ (right).
The top row of spectra are calculated for a pulsar wind with an initial
Lorentz factor $\gammawind(0)=10^3$.
Each subsequent set of three spectra down the page has $\gammawind(0)$
increased by a factor of 10, so that the bottom row represents the emission for
$\gammawind(0)=10^7$.
The other parameters are the same for all panels, and are appropriate for
\PSR\ at periastron.
The dashed curves show the spectra calculated by ignoring the absorption
due to pair production, i.e.\ by artificially setting $\diff\tau/\diff\hat{r}=0$
in Eq.~(\protect{\ref{transport}}).
The solid curves include the effects of the pair production optical depth.
Note that the energy scales are the same for each set of three spectra
for a given wind Lorentz factor, but the scale shifts up a
decade with each factor of ten increase in $\gammawind(0)$.
The energy flux scales are the same for all panels.}
\label{spectra}
\end{figure}

The dependence of the emitted spectra on $\gammawind(0)$ is also striking.
The energy of the scattered photons increases strongly as
the wind Lorentz factor increases; e.g.\ for $\gammawind(0)=10^3$ the peak
of the emitted spectrum occurs just above $10\,\rm MeV$ while for
$\gammawind(0)=10^4$ the peak is at $\approx1000\,\rm MeV$.
The energy of the scattered photons therefore scales as $\sim\gamma^2$
in this regime, as expected since for these Lorentz factors
$\gamma\epsilon_0\ll 1$ and the scattering is in the Thompson regime.
As the wind Lorentz factor is increased further the peak of the spectrum
of scattered photons continues to move up in energy, but the increase
is slower than $\gamma^2$ due to the increasing importance of the
Klein-Nishina effects.
The broadening of the spectrum as $\gammawind(0)$ increases to around $10^5$
and the subsequent narrowing as $\gamma$ increases further is dramatically
illustrated by the left-hand panels for $\thetapulsar=5^\circ$.
This is another example of the effects of tuning the system to maximize the
scattering efficiency, resulting in the greatest deceleration and correspondingly
broad energy distribution of the scattering particles, at $\gamma\epsilon_0\sim1$.

The shape of each individual spectrum is also indicative of the degree of
deceleration effected by the scattering.
All the spectra cut off abruptly at high energies reflecting the maximum energy
$\epsoutplus$ of the scattered photons.
The spectrum calculated for $\gammawind=10^5$ and $\thetapulsar=5^\circ$
rises relatively steeply at the lowest energies shown, then turns over and
rises more slowly for about 1.5 decades in energy (above $10^3\,\rm MeV$)
before dropping steeply towards the upper cutoff.
The steeply-rising portion of the spectrum at low energies is increasing as
$\epsout^{1/2}$ corresponding to a flux density which decreases as
$\epsout^{-1/2}$ -- the usual spectrum for inverse Compton emission from a
single electron in the classical Thompson regime.
This portion of the emitted spectrum is dominated by the electrons at the
minimum Lorentz factor attained after deceleration of the wind.
The flatter portion of the spectrum corresponds to the emission
from the decelerating portion of the wind
--- integrated over a relatively broad range of $\gammawind$;
it is only evident when both $\gammawind(0)$ and
$\thetapulsar$ have values which lead to a substantial deceleration.
When this is not the case the spectra take one of two forms:
if $\gamma\epsilon_0 \ll 1$ the SED increases as $\epsout^{1/2}$ right up to
the cutoff at $\epsoutplus$;
if $\gamma\epsilon_0 \gesim 1$ the Klein-Nishina effects are important
and the SED rises much more steeply to a peak near $\epsoutplus$
and the emission is essentially monochromatic.

Finally, Fig.~\ref{spectra} shows the effects of absorption due to pair production.
For $\gammawind(0)\lesim10^5$ the scattered photons are below the threshold energy
required for pair production on the target photons even when $\thetapulsar$ is small,
and so there is no absorption.
The absorption due to pair production is most significant in the spectrum plotted for
$\gammawind(0)=10^6$ and $\thetapulsar=5^\circ$ where it removes a substantial part
of the emission at energies up to a decade below the peak.
For smaller $\gammawind$ the scattered photons are below the threshold for absorption,
whereas at higher values the resonant character of the cross section becomes apparent.

\subsection{Light curves}
The inclination angle of the orbit of \PSR\ is $i=35^\circ$ so a
telescope on the Earth will sample angles $\thetapulsar$ between $90-i=55^\circ$ and
$90+i=125^\circ$ over the binary period.
These extremes are the two larger values of $\thetapulsar$ for which the spectra are
plotted in Fig.~\ref{spectra}.
The increase from the minimum value of $\thetapulsar$ to
the maximum occurs over just 65 days beginning on day $-6$, just before periastron.
The subsequent decrease occurs far more slowly over the remaining 1170 days
of the orbital period.
This variation, together with the variation in the target photon density due to the
changing binary separation during the orbit, results in a profound variation in the
emission of inverse Compton scattered radiation from the binary system.

\begin{figure}[htb]
\epsfxsize=14cm\epsffile{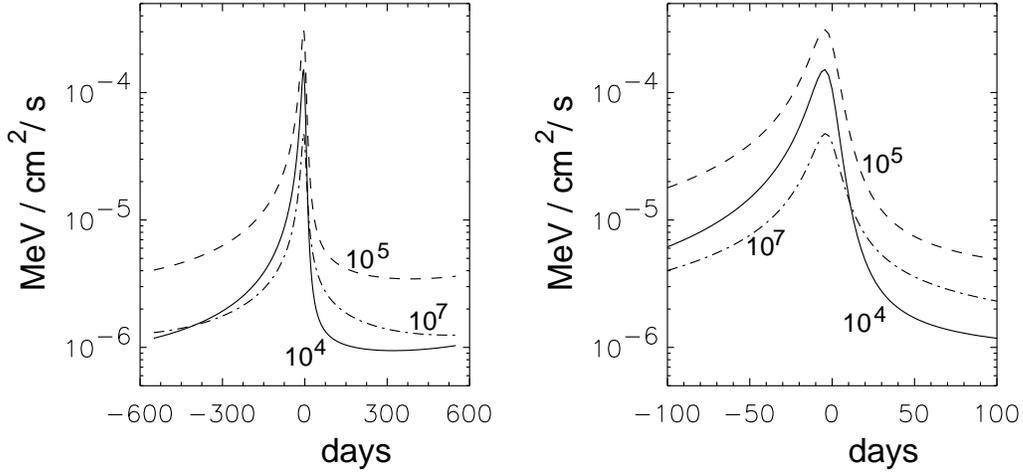}
\caption{Light curves showing the integrated energy flux at the Earth from
\PSR . The different curves correspond to different initial wind
speeds: the solid curve is for $\gammawind(0)=10^4$,
the dashed curve is for $\gammawind(0)=10^5$, and the dot-dashed
curve is for $\gammawind(0)=10^7$.
The left panel shows the variation over the entire orbit
and the right panel shows just 200 days centered on periastron.}
\label{lightcve}
\end{figure}

Fig.~\ref{lightcve} shows the orbital variation of the integrated flux density,
$\int F_E \,\diff E$, expected from \PSR\ for $\gammawind(0)=10^4$, $10^5$ and $10^7$.
The light curve for $\gammawind(0)=10^6$ is very similar to that for $\gammawind(0)=10^5$
so it has been omitted for clarity.
The total variation in the integrated flux is very large;
for $\gammawind=10^4$ the peak is two orders of magnitude above the minimum.
The light curves are also very strongly peaked and clearly asymmetric,
with the increase towards the maximum integrated
flux occurring somewhat more gradually than the decline to the minimum.
The asymmetry, peak to minimum ratio, and peaked nature of the light curves are
minimized when $\gammawind(0)\epsilon_0\sim 1$, but even then the
peak to minimum ratio is more than a factor of ten.
These features are all a direct reflection of the orbital variation of
the binary separation and the angle $\thetapulsar$;
the asymmetry is a straightforward manifestation of the asymmetry in the
variation of $\thetapulsar$ about periastron.

The peak of the spectrum of the scattered photons also varies with orbital phase.
If this variation were to cross the threshold of an observing instrument the
detected light curve could be somewhat different from those shown in Fig.~\ref{lightcve}.
The energy spectra at the extreme values of $\thetapulsar$ sampled during one
binary orbit are shown in the second and third columns of Fig.~\ref{spectra}.
The variation in the peak energy is largest when $\gammawind(0)\epsilon_0\lesim 1$.
For $\gammawind(0)=10^4$ the peak is at $\sim 1000$\,MeV just prior to
periastron (when $\thetapulsar$ attains its minimum value),
and it moves down to $\sim 200$\,MeV by about day +60
(corresponding to the maximum of $\thetapulsar$).
For $\gammawind(0)=10^5$ the peak moves from $\sim 30$\,GeV to $\sim 20$\,GeV,
while for $\gammawind(0)=10^6$ and $\gammawind(0)=10^7$ the peak is
at $\sim 300$\,GeV and $\sim 5$\,TeV respectively, and moves very little.

The integrated flux shown in Fig.~\ref{lightcve} does not include the effects
of absorption due to pair production.
Calculations including the pair production optical depth indicate that its
effect on the integrated flux density from \PSR\ is small -- reducing the
integrated flux density by $\sim10$\% or less -- and that the reduction
is relatively insensitive to the orbital phase.
It is therefore unimportant for observations of the inverse Compton emission.

\newpage

\section{Discussion}

The models for the inverse Compton emission from the freely-expanding wind
of \PSR\ presented in \S 5 suggest that the emission is likely to be detectable
at energies somewhere in the GeV to TeV range, depending on the wind parameters.
Given the present lack of operational $\gamma$-ray telescopes,
the best opportunity for detecting the emission is in the TeV energy range.

In calculating the inverse Compton emission spectra we optimistically assume
that none of the wind momentum is carried by ions and that
the wind is not terminated via pressure balance with the companion star outflow.
The results indicate that if the initial Lorentz factor of the wind
is in the range $10^6-10^7$ the inverse Compton emission from
the unshocked wind at epochs near periastron could be as much as a factor of
$100-1000$ above the sensitivity threshold of the new CANGAROO II
imaging Cerenkov detector
[Yoshikoshi et al.\ 1999]
which should be operational before the end of 1999.
If the pulsar wind is slower, the inverse Compton emission should
be detectable by the GLAST observatory at energies in the range 20\,MeV--300\,GeV
if $\gammawind(0)\sim 10^4-10^5$,
or by the INTEGRAL project at energies up to 10\,MeV if $\gammawind(0)\sim 10^3$.
The sensitivity of each of these planned satellite-borne observatories
is easily sufficient to detect the model emission.
However, they will not operate until well after the next periastron.

If either of the fundamental assumptions regarding the pulsar wind
is not applicable to the \PSR\ system,
the gamma-ray emission from the unshocked wind will be lower than
suggested by the models detailed in \S 5.
If the momentum flux of the pulsar wind is sufficiently small that it
is dominated by the Be-star outflow, then the pulsar wind will terminate
close to the pulsar at a shock which is wrapped around the pulsar.
The mismatch between the shape of this termination shock and the loss
surfaces due to inverse Compton scattering could then lead to
integrated-flux light curves that are quite different to those
presented in Fig.~\ref{lightcve}.
Furthermore, the inverse Compton deceleration of the pulsar wind
decreases its ram pressure, reducing the termination shock radius,
but this effect will be small at values of $\thetapulsar$ which
can be observed.
The nature of the gamma-ray light curves will depend on both the
stand-off distance and the details of the shape and orientation
of the pulsar wind termination shock.

Observations of changes in the pulsed emission from \PSR\
[Johnston et al.\ 1996]
and observations and interpretation of the unpulsed radio emission
seen around periastron
[Johnston et al.\ 1999; Ball et al.\ 1999]
indicate that the pulsar and its wind interact with a disk surrounding
the Be star between about day $-25$ and $+25$.
There are few observations of X-ray and $\gamma$-ray emission from this
system at epochs outside this range.
Tavani \& Arons [1997]
have addressed the problem of fitting the X-ray
observations close to periastron with the inverse Compton emission from
the shocked pulsar wind.
Their analysis indicates that the details of the flow pattern,
the competition between adiabatic and radiative losses,
and strong shielding (or shadowing) of the target photons by the pulsar disk,
are all important near the epoch of periastron.
These details are not considered in our work here, and the light curves of
Fig.~\ref{lightcve} can therefore not be expected to be accurate
at epochs close to periastron.
Nevertheless the upper limits from EGRET measurements
[Tavani et al.\ 1996]
just prior to the 1994 periastron
place some constraint on the pulsar wind parameters.
In particular, the inverse Compton spectrum from an unbounded lepton wind
with an initial Lorentz factor of $\gammawind(0)=10^4$
(as shown in Fig.~\ref{spectra})
would exceed the EGRET upper limits which imply that the energy flux
of photons at energies of 30\,MeV to 10\,GeV is less than
about $2\times10^{-5}\,{\rm MeV\,cm^{-2}\,s^{-1}}$.

The periastron to apastron ratio of inverse Compton emission from an
unbounded lepton wind from \PSR\ as shown in Fig.~\ref{lightcve}
is much larger than the flux ratio from the shock-accelerated electrons
when the wind terminates close to the pulsar
[Kirk et al.\ 1999].
Observations of the high-energy light curve may therefore enable the deconvolution
of the two contributions to the $\gamma$-ray emission.
In particular, observations aimed at detecting an asymmetric light curve at energies
in the MeV--TeV range, or at least placing firm upper limits on the
high energy emission from this system, should provide valuable information
on the properties of the unshocked pulsar wind.

\begin{ack}
We thank Simon Johnston for many helpful discussions,
and LTB thanks Olaf Skj\ae raasen for assistance with
Fig.\ \ref{shockgeom}.
Our collaboration was made possible by support for
JGK by the RCfTA under its International Visitor program,
and for LTB from the MPIK Heidelberg.
\end{ack}

\newpage

\section*{Appendix A: Energy and momentum balance}
Assuming the magnetic field to be weak,
the equations of conservation of energy and radial momentum flux in the
wind take on the forms:  
\eqb {1\over r^2}{\diff
\over\diff r}\left[r^2 \left(E+P\right)\betawind\gammawind^2\right] 
&=& C^0\,\equiv\,
n_{\rm e} \left<{1\over c}{\diff e \over \diff \tau}\right>
\label{energy}\\ 
{1\over r^2}{\diff \over\diff r}
\left\{ r^2\left[ \left(E+P\right)
\betawind^2\gammawind^2 + P\right]\right\} 
&=& C^r\,\equiv\, n_{\rm e} \left<{\diff p_r \over \diff \tau}\right>
\label{momentum} 
\eqe 
[e.g., Landau \& Lifshitz 1959]
Here, $E$ and
$P$ denote the proper gas energy density (including rest-mass) and pressure,
respectively.
The radial velocity of the gas is $c\betawind$ and the associated Lorentz
factor is $\gammawind$.
The components $C^0$ and $C^r$ of the four-force
density on the right-hand sides of Eqs.~(\ref{energy}) and (\ref{momentum})
are proportional to the proper number density $n_{\rm e}$ of 
electrons and positrons taken together and to 
the average rates of energy and radial momentum change of a single electron
due to Compton scattering:
$\left<\diff e/\diff\tau\right>$ and $\left<\diff p_r/\diff\tau\right>$ 
where $\tau$ is the proper time, measured in the rest frame of the wind.

The average rate of change of energy of the photon field is given by
an integral similar to that in equation (\ref{scattrate}) with the
additional weight factor ${\epsout}-{\epsin}$.
Equating the energy loss rate by the electron to the gain by the
photon field leads, after a straightforward calculation, to the expression
\eqb
\left<{1\over c}{\diff e\over\diff t}\right>
&=& -
{3\over8}\sigmaT \me c^2
\int
\diff\vec{\Omega}\diff{\epsin} 
{{{\epsinp}}^2\over{\epsin}}
\left(1-{{\epsin}\over\gamma{\epsinp}}-{{\epsin}\over\gamma}\right)\times\nonumber\\
&&n_\gamma({\epsin},{\bf R},\vec{\Omega}) F_{\rm loss}({\epsinp})\;.
\label{enexch}
\eqe

The rate of change of radial momentum is calculated in a similar 
fashion, using the quantity
$-{\epsout}\cos\thetaout+{\epsin}\cos\thetain$ as the weighting function,
which gives
\eqb
\left<{\diff p_r\over\diff t}\right>
&=& -
{3\over8\beta}\sigmaT \me c\int\diff\vec{\Omega}\diff{\epsin} 
{{{\epsinp}}^2\over{\epsin}}
\left(1-{{\epsin}\over\gamma{\epsinp}}-
{{\epsin}\over\gamma}+{{\epsinp}\over\gamma^2}\right)\times
\nonumber\\
&&n_\gamma({\epsin},{\bf R},\vec{\Omega}) F_{\rm loss}({\epsinp})\;.
\label{momexch}
\eqe

The amount of heating is given by the projection of the four-force density
onto the velocity.
Inspection of Eqs.~(\ref{enexch}) and (\ref{momexch}) reveals that 
only the fourth term ${\epsinp}/\gamma^2$ in parentheses in
Eq.~(\ref{momexch}) survives the projection.
In the case of an ultra-relativistic wind this term is small 
($\sim{\epsin}/\gamma$) compared to the leading ones 
in the expression for the work done decelerating the electron,
and we  neglect it in the following.  
In this approximation, we set $P=0$ and
$E/c^2=\rhowind$, the proper rest-mass density,
and identify the Lorentz
factor  of the wind with that of each individual electron.
As a result, one of the equations (\ref{energy}) and (\ref{momentum})
is superfluous.
Keeping the energy equation (\ref{energy}), and combining it with the
equation of mass conservation, which reads
\eqb
{1\over r^2}{\diff \over\diff r}\left(
r^2\rhowind\betawind\gammawind\right) 
&=& 0 
\eqe 
one finds 
\eqb 
{\diff\gammawind
\over\diff r} 
&=& {\eta\over \me c^3\betawind\gammawind}\left<{\diff e\over \diff\tau}\right>
\nonumber\\
&=& {\eta\over \me c^3\betawind}\left<{\diff e\over\diff t}\right>\;.
\label{dgamma}
\eqe
Equation (\ref{energy2}) follows directly through substitution of
(\ref{enexch}) into (\ref{dgamma}).
\newpage

\section*{Appendix B: Radiation transfer of inverse Compton emission }
The differential number density of electrons in the pulsar wind
follows directly from equation (\ref{windees}) and is given by
\eqb
{\diff N_{\rm e}\over\diff\gamma\diff\vec{\Omega}\diff r}&=&r^2 N_{\rm 
e}(r)\delta\left[\gamma-\gammawind(r)\right]
\nonumber\\
&\approx&{\eta\Lwind\over 4\pi\gammawind(0)\me c^3}\delta[\gamma-\gammawind(r)]
\eqe
where we have assumed $\betawind\approx1$.

The radiative transport equation governing the specific intensity
$I({\epsout},r)$, including both emission by inverse Compton 
scattering and absorption by photon-photon pair production, is
\eqb
{\diff I\over\diff r}&=& \epsout \me c^2\eta_{\rm ic} + I{\diff\tau\over\diff r}\;.
\eqe
The quantity $\eta_{\rm ic}$ is the inverse Compton emissivity,
the number of photons emitted per second per 
unit volume per unit energy interval around $\epsout \me c^2$,
which is given by
\eqb
\eta_{\rm ic}&=& 
{1\over  \me c^2}\int\diff\gamma
 {\diff N_{\rm e}\over\diff\gamma\diff\vec{\Omega}\diff r}
 {\diff N_{\gamma}\over\diff{\epsout}\diff t}
 \nonumber\\
 &=&\left({\Lwind\over4\pi}\right){\eta\over 
\gammawind(0)(\me c^2)^2 c} 
 {\diff N_{\gamma}\over\diff{\epsout}\diff t}
 \eqe
where $\diff N_{\gamma}/\diff{\epsout}\diff t$ 
is the differential rate of emission of photons by a single electron.

The function ${\cal H}({\epsout},\cos\psi)$
which appears in the photon production rate,
Eq.~(\ref{photonproduction}),
is related to the quantity
${\diff N_\gamma/\diff t\diff{\epsout}[\diff{\epsin}\diff x']}$
derived by Ho \& Epstein~[1989, Eq.~A14],
and is defined by
\eqb
{\cal H}({\epsout},\cos\psi)&=&
{3\xi\over8\gamma^2\epsin} \;
{1\over[(1-\lambda-\xi)^2+\beta^2(1-x'^2)]^{1/2}}\times \nonumber\\
&&
\left[
{1\over \xi} + {y_0^2\over (a^2-b^2)^{1/2}} +
\left({2y_0\over\varepsilon} + {a\over\varepsilon^2}\right)+
\right.\nonumber\\
&&
\left.\left({a\over(a^2-b^2)^{1/2}}-1\right)
+ {a\xi-a^2+b^2\over(a^2-b^2)^{3/2}}
\right]
\eqe
with
\eqb
\xi=1-\betawind x', \qquad \lambda=\epsout/\gammawind,
\qquad \varepsilon=\epsin/\gammawind,
\eqe
\eqb
x'={\betawind-\cos\psi\over1-\betawind\cos\psi}
\eqe
and where
\eqb
a=\xi+\varepsilon(1-y_0), \qquad\qquad b=-\varepsilon\delta
\eqe
\eqb
y_0=
{(1-\lambda-\xi)(1-\lambda-\rho\xi)\over
(1-\lambda-\xi)^2+\beta^2(1-x'^2)}\;,
\eqe
\eqb
\delta=
{\beta(1-x'^2)^{1/2}[\beta^2+2\lambda(1-\rho)\xi-(1-\rho\xi)^2]^{1/2}
\over
(1-\lambda-\xi)^2+\beta^2(1-x'^2)}\;,
\eqe
and finally
\eqb
\rho={\epsout\over\epsin}\;.
\eqe
However, this function gives the photon production rate
only when $\epsoutminus \leq \epsout \leq \epsoutplus$
where
\eqb
\epsoutminus =
{\gamma\varepsilon(1-\beta\cos\psi)
\over
(1+\varepsilon+Q_\varepsilon)}\;,\qquad{\rm and}
\qquad
\epsoutplus =
{\gamma\varepsilon(1+\varepsilon+Q_\varepsilon)
\over
(\xi+2\varepsilon)}
\eqe
with
\eqb
Q_\varepsilon=(\beta^2+\varepsilon^2+2\beta\varepsilon\cos\psi).
\eqe
At these limits the term
$[\beta^2+2\lambda(1-\rho)\xi-(1-\rho\xi)^2]^{1/2}$
in the quantity $\delta$ becomes zero, and outside this range the photon
production rate is zero.

The differential optical depth due to pair production was given by 
Kirk et al.~[1999, Eq.~B4]
In our notation it is
\eqb
\hat{\sigma}_{\gamma\gamma}(x)=
\left\{
\begin{array}{ll}
{3\over 16} (1-\zeta^2)
\left[
(3-\zeta^4)\ln\left({1+\zeta\over1-\zeta}\right)
-2\zeta(2-\zeta^2)
\right] & ~~~{\rm if} ~x\geq 1 \\
0 & ~~~{\rm otherwise}
\end{array}\right.
\eqe
where 
\eqb
\zeta={(x^2-1)^{1/2}\over x}\;.
\eqe

\clearpage

{}

\begin{thebibliography}{}

\bibitem{aharonianbogovalov99}
Aharonian, F. A., Bogovalov, S. V., 1999,
Astron. Nach., in press

\bibitem{aschenbachbrinkmann75}
Aschenbach, B., Brinkmann, W., 1975,
A\&A, 41, 147

\bibitem{balletal99}
Ball, L., Melatos, A. M., Johnston, S., Skj\ae raasen, O., 1999,
ApJ, 541, L39

\bibitem{belletal95}
Bell, J. F., Bessell, M. S., Stappers, B. W., Bailes, M., Kaspi, V. M., 1995,
ApJ, 447, L117

\bibitem{bogovalovaharonian99}
Bogovalov, S. V., Aharonian, F. A., 1999,
MNRAS, submitted

\bibitem{CI99}
Chernyakova, M. A., Illarionov, A. F., 1999,
MNRAS, 304, 359

\bibitem{dejagerharding92}
De Jager, O. C., Harding, A. K., 1992,
ApJ, 396, 161

\bibitem{emmeringchevalier87}
Emmering R. T., Chevalier R. A., 1987, 
ApJ, 321, 334

\bibitem{harding96}
Harding, A. K., 1996,
Space Sc.\ Rev., 75, 257

\bibitem{hesteretal95}
Hester, J. J., 1995,
ApJ, 448, 240

\bibitem{hirayamaetal96}
Hirayama, M., Nagase, F., Tavani, M., Kaspi, V. M., Kawai, N., Arons, J., 1996,
PASJ, 48, 833

\bibitem{hoepstein89}
Ho, C., Epstein, R. I., 1989,
ApJ, 343, 277

\bibitem{johnstonetal92}
Johnston, S., Manchester, R. N., Lyne, A. G., Bailes, M., Kaspi, V. M.,
Qiao, G., D'Amico, N., 1992,
ApJ, 387, L37

\bibitem{johnstonetal94}
Johnston, S., Manchester, R. N., Lyne, A. G.,
Nicastro, L., Spyromilio, J., 1994,
MNRAS, 268, 430

\bibitem{johnstonetal96}
Johnston, S., Manchester, R. N., Lyne, A. G., D'Amico, N.,
Bailes, M., Gaensler, B. M., Nicastro, L., 1996,
MNRAS, 279, 1026

\bibitem{johnstonetal99}
Johnston, S., Manchester, R. N., McConnell, D., Campbell-Wilson, D., 1999,
MNRAS, 302, 277

\bibitem{jones65}
Jones, F. C., 1965,
Phys.\ Rev.\ B, 137, 1306

\bibitem{kaspietal94a}
Kaspi, V. M., Johnston, S., Bell, J. F. Manchester, R. N.,
Bailes, M., Bessell, M., Lyne, A. G., D'Amico, N., 1994a,
ApJ, 423, L43

\bibitem{kaspietal94b}
Kaspi, V. M., Bell, J. F., Bessell, M., Stappers, B.,
Bailes, M., Manchester, R. N., 1994b,
AAS, 185, 6207

\bibitem{kennelcoroniti84}
Kennel, C. F., Coroniti, F. V., 1984,
ApJ, 283, 710

\bibitem{KBS99}
Kirk, J. G., Ball, L., Skj\ae raasen, O., 1999,
Astroparticle Phys., 10, 31

\bibitem{ll59}
Landau, L. D., Lifshitz, E. M., 1959,
Fluid Dynamics, Pergamon press: Oxford

\bibitem{mastichiadis91}
Mastichiadis, A., 1991,
MNRAS, 253, 235

\bibitem{melatosetal95}
Melatos, A., Johnston, S., Melrose, D. B., 1995,
MNRAS, 275, 381

\bibitem{qiaolin1998}
Qiao, G. J., Lin, W. P., 1998,
A\&A, 333,172

\bibitem{luoprotheroe98}
Luo, Q., Protheroe, R. J., 1998,
Publ.\ Astron.\ Soc.\ Aust., 15, 222

\bibitem{reesgunn74}
Rees, M. J., Gunn, J. E., 1974,
MNRAS, 167, 1

\bibitem{sakoetal97}
Sako, T., et al., 1997,
Proc.\ 25th ICRC (Durban), 3, 165

\bibitem{tavanietal96}
Tavani, M., et al., 1996,
A\&AS, 120, 221

\bibitem{tavaniarons97}
Tavani, M., Arons, J., 1997,
ApJ, 477, 439

\bibitem{underhilldoazan82}
Underhill, A., Doazan, V., 1982,
`B stars with and without emission lines', NASA SP-456

\bibitem{yoshikoshi99}
Yoshikoshi, T., et al., 1999,
Astroparticle Phys., 11, 267

\bibitem{zhangetal97}
Zhang, B., Qiao, G. J., Han, J. L., 1997,
ApJ, 491, 891
 
\end{thebibliography}
\end{document}